\documentclass[twocolumn,a4paper,10pt]{article}
\usepackage{amsmath,amsfonts}
\usepackage{epsfig}
\usepackage{times}
\usepackage{epsfig}
\usepackage[small,hang]{caption}

\usepackage{graphicx}
\usepackage{subcaption}
\usepackage{multirow}

\makeatletter  

\newcounter{numbersec}
\renewcommand{\section}[1]{\par\noindent\stepcounter{numbersec}
	\par
	\vspace{6pt}
	\noindent\textbf{\large   \arabic{numbersec} \hspace*{0.3cm} #1 }
	\par
	\vspace{2pt}
}
\renewcommand{\subsection}[1]{
	\par
	\vspace{6pt}
	\noindent\textbf{#1}
	\par
}
\renewcommand{\subsubsection}[1]{%
	\par
	\vspace{6pt}
	\textbf{#1.}
}

\newcommand{\Abstract}{\par\vspace{6pt}\noindent\textbf{\large Abstract}\par\vspace{2pt}}
\newcommand{\Acknowledgments}{\par\vspace{6pt}\noindent\textbf{\large Acknowledgments }\par\vspace{2pt}}

\newenvironment{References}{
\par\vspace{6pt}\noindent\textbf{\large References}\par\vspace{2pt}
\begin{small}\begin{list}{ }{\itemsep2mm \parsep0mm\labelsep0mm\leftmargin0mm}}
{\end{list}\end{small}}

\makeatother

\title{\vspace*{-12mm}
\LARGE \sc \textbf{  
Discovering Flow Separation Control Strategies in 3D Wings via Deep Reinforcement Learning
}}

\author{ \Large \bf \textit{ 
R. Montalà$^{1}$, B. Font$^{2}$, P. Suárez$^{3}$, J. Rabault$^{4}$, O. Lehmkuhl$^{5}$,} \\
\Large \bf  \textit{ R. Vinuesa$^{6}$ and I. Rodriguez$^{1}$}  \\ \\
\bf  $^{1}$ \textit{ TUAREG, Universitat Politècnica de Catalunya (Spain)} \\
\bf  $^{2}$ \textit{ Mechanical Engineering, Delft University of Technology (Netherlands)} \\
\bf  $^{3}$ \textit{ FLOW, Engineering Mechanics, KTH Royal Institute of Technology (Sweden)} \\
\bf  $^{4}$ \textit{ Independent Researcher (Norway)} \\ 
\bf  $^{5}$ \textit{ LS/CFD - CASE, Barcelona Supercomputing Center (Spain)} \\
\bf  $^{6}$ \textit{ Department of Aerospace Engineering, University of Michigan (USA)} \\
{\it ricard.montala@upc.edu}
}
\date{}

\begin{document}
\maketitle
\thispagestyle{empty}



\Abstract

In this work, deep reinforcement learning (DRL) is applied to active flow control (AFC) over a three-dimensional SD7003 wing at a Reynolds number of $Re=60{,}000$ and angle of attack of $\text{AoA}=14^\circ$. In the uncontrolled baseline case, the flow exhibits massive separation and a fully turbulent wake. Using a GPU-accelerated CFD solver and multi-agent training, DRL discovers control strategies that enhance lift ($79\%$), reduce drag ($65\%$), and improve aerodynamic efficiency ($408\%$). Flow visualizations confirm reattachment of the separated shear layer, demonstrating the potential of DRL for complex and turbulent flows.


\section{Introduction}
Active Flow Control (AFC) has gained significant attention in recent years as a promising strategy for improving aerodynamic efficiency and reducing environmental impact in the transportation sector. By manipulating the flow around bluff bodies, such as road vehicles (Cerutti et al., 2020), trucks (Minelli et al., 2016), or airfoils at high angles of attack (Rodriguez et al., 2020), AFC techniques can reduce drag, leading to improved fuel efficiency and lower CO$_2$ emissions. Traditional AFC methods rely on predefined actuation strategies such as steady blowing, suction, or harmonic forcing, often based on empirical tuning and fixed parameters. In contrast, deep reinforcement learning (DRL) offers a fundamentally different and potentially more powerful approach. Thanks to the recent breakthroughs in machine learning (ML), DRL appears particularly well-suited for flow control problems (Garnier et al., 2021), enabling data-driven and closed-loop strategies in which the control policy is learned directly through interaction with the flow. Unlike conventional approaches, DRL can uncover complex, time-dependent relationships between actuation inputs and flow responses. This allows for dynamic control strategies that evolve over time and adapt to instantaneous flow conditions, often yielding superior performance compared to fixed-frequency or open-loop methods.

The application of DRL to active flow control has gained traction in recent years, beginning with the pioneering work of Rabault et al. (2019), who achieved drag reduction on a two-dimensional cylinder at a Reynolds number of $Re = 100$. Building on this work, Rabault and Kuhnle (2019) introduced multi-environment training strategies to improve learning efficiency and scalability to higher Reynolds numbers. More recently, Suárez et al. (2025a, 2025b) advanced the methodology further by applying DRL to three-dimensional cylinders, achieving notable drag reductions even at $Re = 3,900$. Beyond bluff body control, DRL has also been employed to reduce skin friction in wall-bounded turbulence at $Re_\tau = 180$ (Guastoni et al., 2023) and to three-dimensional Rayleigh–Bénard problems (Vasanth et al., 2024). However, in the context of airfoils and wings, DRL applications remain limited, with existing studies largely focused on two-dimensional geometries operating at low Reynolds numbers (Wang et al., 2022 and Garcia et al., 2025).

Despite these advancements, the combination of DRL and AFC remains in an early stage of development. Most existing applications are restricted to low Reynolds numbers and/or canonical geometries, limiting their direct applicability to industrial scenarios. To address this gap, the present study introduces an AFC-DRL framework that leverages a GPU-accelerated computational fluid dynamics (CFD) solver for efficient data generation during training, thereby enabling the extension of DRL-based control strategies to more complex and realistic cases. This framework has been previously validated in various flow control scenarios, including a turbulent boundary layer separation at $Re_\tau = 180$ (Font et al., 2025), flow control over a three-dimensional cylinder at $Re = 100$ (Montalà et al., 2024), and over a three-dimensional wing at $Re=1,000$ (Montalà et al., 2025). In this work, we extend the applicability of AFC-DRL to a more realistic aerodynamic configuration: flow over a three-dimensional SD7003 wing at a Reynolds number of $Re = 60{,}000$ and an angle of attack of $\text{AoA}=14^\circ$. This case is characterised by massive flow separation over the wing and a fully turbulent wake, making it a challenging benchmark for flow control. To the best of our knowledge, this represents the most complex case to date combining DRL and AFC, featuring both a moderately complex geometry and fully turbulent flow conditions at a moderate Reynolds number.

\section{Methodology}

\begin{figure}[h]
	\begin{center}
	\includegraphics*[width=0.9\linewidth]{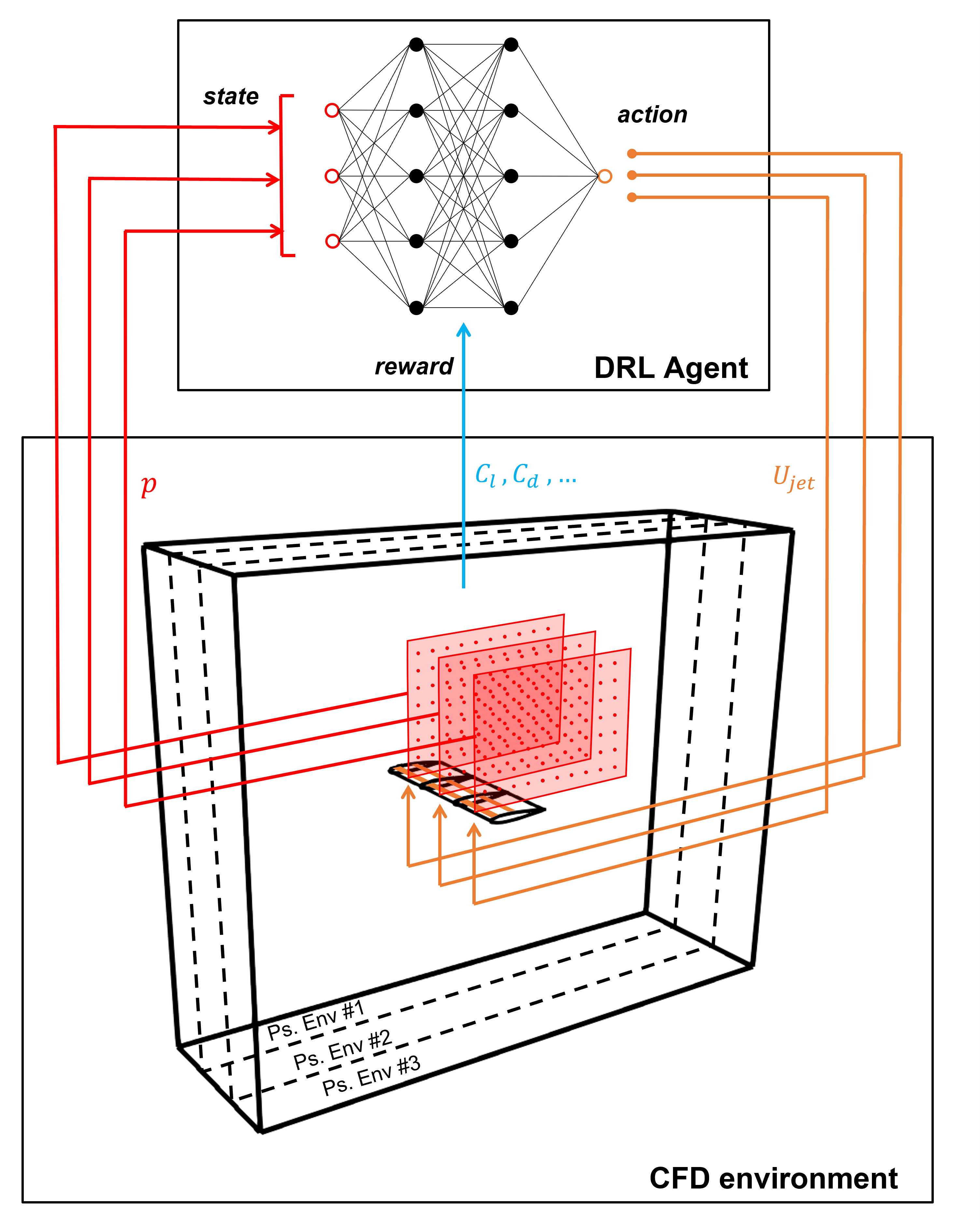}
	\caption{\label{fig:DRL_framework} CFD-DRL set-up.}
	\end{center}
\end{figure}

As illustrated in Fig. \ref{fig:DRL_framework}, the DRL framework consists of two main components: The CFD environment and the DRL agent. The agent is a neuronal network (NN) that maps flow states to corresponding control actions. In this setup, the NN receives pressure signals from flow sensors within the CFD environment and outputs the jet velocity to be applied back into the system. To optimise the control strategies, the agent undergoes a training process based on trial-and-error learning. Therefore, the CFD solver and the agent not only exchange states and actions but also a reward. The reward represents the quantity to be optimised, typically comprising variables that characterise aerodynamic performance, such as the lift or drag coefficient.

\subsection{Environment Configuration}
The environment is modelled using the SOD2D solver (Gasparino et al., 2024), a GPU-enabled spectral element method (SEM) code developed at the Barcelona Supercomputing Center (BSC). Its scalability and computational efficiency on distributed accelerator-based architectures make it well-suited for DRL applications, where rapid CFD data generation is crucial for the training. In the present work, the filtered incompressible Navier-Stokes equations (LES) are solved. These equations are written as follows:

\vskip-.4cm
\begin{eqnarray}
\label{eq:NS_1} \frac{\partial \tilde{u}_i}{\partial x_i} = 0 \; , \\
\label{eq:NS_2} \frac{\partial \tilde{u}_i}{\partial t} + \tilde{u}_j \frac{\partial \tilde{u}_i}{\partial x_j} = -\frac{1}{\rho} \frac{\partial \tilde{p}}{\partial x_i} + \nu \frac{\partial^2 \tilde{u}_i}{\partial x_j^2} - \frac{\partial \tau_{ij}}{\partial x_j} \; ,
\end{eqnarray}

\noindent where $\tilde{u}_i$ and $\tilde{p}$ denote the filtered (resolved) velocity and pressure fields, $\rho$ the fluid density, $\nu$ is the kinematic viscosity and $\tau_{ij}$ the unresolved subfilter stress (SFS) tensor. In this case, the SFS tensor is evaluated using the ILSA model (Piomelli et al., 2015).

The computational domain consists of a box with dimensions $L_x = 19c$, $L_y = 20c$ and $L_z=3.0c$ along the streamwise, crosswise and spanwise directions, respectively. The wing is positioned approximately in the centre and extends the entire domain width in the spanwise direction. Periodic boundary conditions are applied in this direction, thereby modelling an infinite wing. A uniform velocity $[u,v,w] = U_\infty \, [\text{cos} \, (\text{AoA}), \text{sin} \, (\text{AoA}), 0]$ is prescribed at the inlet, with the corresponding Reynolds number $Re=60{,}000$ and angle of attack $\text{AoA}=14^\circ$. At the outlet, zero-gradient conditions are applied for velocity, along with a constant pressure condition. A no-slip boundary condition is imposed on the wing surface, while Dirichlet boundary conditions are applied at the jet surfaces to prescribe the corresponding jet velocity when AFC is activated.

To explore three-dimensional actuations without increasing the computational cost of training, a multi-agent reinforcement learning (MARL) framework is employed (Belus et al., 2019). This approach has also been applied by Suárez et al. (2025a, 2025b), Montalà et al. (2024, 2025) and Font et al. (2025). The method leverages the spanwise invariance of the problem by dividing the domain into multiple subdomains (or pseudo-environments), each equipped with its own jet actuator. All pseudo-environments share a single agent that, based on the local state vector provided by each pseudo-environment, outputs distinct jet velocities. This enables spatially distributed actuation while maintaining a manageable training complexity. In the present work, three MARL pseudo-environments are employed ($n_\text{MARL}=3$). Furthermore, to enhance parallel data collection and reduce wall-clock time, ten independent CFD simulations are run concurrently ($n_\text{CFD}=10$) during training (Rabault and Kuhnle, 2019). Consequently, at the end of each episode, a total of $3 \times 10 = 30$ state-action-reward trajectories are collected to perform the policy update. The duration of an episode corresponds to $T_\text{eps} = 10.52 U_\infty/c$. Within each episode, 120 actions are applied per MARL pseudo-environment, meaning that the duration of each action is $T_\text{act} = T_\text{eps}/120$.

Since three MARL pseudo-environments are used, each one spans a subdomain of length  $L_{z,\text{MARL}} = L_z/3$. Within each subdomain, a pair of jets is considered, one near the leading edge of the wing at approximately $x/c = 0.01$, and the other further downstream at $x/c = 0.4$. Both jets extend across the entire span of their corresponding subdomain. The agent controls the front jet, while the downstream jet is constrained to have the opposite instantaneous velocity, thereby ensuring mass conservation at each time step. For the state, each pseudo-environment includes its own slice of 90 sensors in its z-midplane. To provide three-dimensional flow information, the input of the NN has a size of $90 \times 3 = 270$, incorporating data from adjacent slices on either side.

\subsection{Agent Configuration}
The DRL agent is implemented using the TF-Agents Python library (Guadarrama et al., 2018) and communicates with the CFD solver through a Redis in-memory database, managed via SmartSim (Partee et al., 2022). This setup effectively resolves the "two-language" problem by enabling low-overhead interaction between the Fortran-based CFD solver and the Python-based DRL agent.

The Proximal Policy Optimisation (PPO) algorithm (Schulman et al., 2017) is employed to refine the agent’s policy over the training episodes. PPO is an actor-critic reinforcement learning method. In this work, both the actor and the critic networks are modelled as multi-layer perceptrons (MLP) comprising two hidden layers of 512 neurons each. The actor network maps observed states to actions (as illustrated in Fig. \ref{fig:DRL_framework}). The critic network, on the other hand, estimates the advantage function, i.e., an assessment of how favourable a particular action is relative to the expected performance under the current policy. This estimation guides the learning process by indicating whether selected actions are better or worse than average. The critic is trained via supervised learning at the end of each episode, using trajectory data collected during interaction with the environment. The actor is subsequently updated using the advantage estimates, ensuring that the policy is adjusted in a direction that improves the expected accumulated reward while maintaining stability, as enforced by PPO’s clipped objective function.

In this work, the training is conducted using the reward function written in Eq. \ref{eq:Reward_global}. Thus, the reward accounts for both local and global performance, with the weighting factor $\gamma$ modulating the relative importance of the local reward $r_i$ versus the mean reward computed across all pseudo-environments.

\vskip-.5cm
\begin{equation} \label{eq:Reward_global}
	R_i = \gamma \; r_i + (1-\gamma)/n_\text{MARL} \sum_{j=1}^{n_\text{MARL}}r_j \; ,
\end{equation}
\begin{equation} \label{eq:reward}
	r_i = (C_{d,b} - C_d) - \alpha \; | C_l - C_{l,avg} | + \beta(C_l - C_{l,b}) \; .
\end{equation}

The local reward $r_i$ is computed as defined in Eq. \ref{eq:reward}, where $C_{d,b}$ and $C_{l,b}$ denote the drag and lift coefficients of the baseline (uncontrolled) scenario, respectively, $C_d$ and $C_l$ represent the instantaneous drag and lift coefficients obtained during the episode, and $C_{l,\text{avg}}$ corresponds to the accumulated mean lift coefficient over the course of the episode. The reward function thus comprises three contributions: the first term encourages a reduction in drag relative to the baseline; the second penalises fluctuations in the lift coefficient, thereby indirectly suppressing vortex-shedding behaviour; and the third promotes an increase in lift compared to the baseline. In the present study, the reward hyperparameters are selected to be $\alpha = 0.3$, $\beta=0.5$ and $\gamma = 0.8$.

\section{Results}

To reduce the computational cost and make training practically feasible, two levels of mesh refinement are employed: a coarse mesh for training the agent and a fine mesh for evaluating the final policy and comparing it with the baseline. Although the coarse mesh does not correctly resolve all flow scales, it effectively captures the large-scale structures that dominate the flow dynamics, thereby providing sufficiently informative feedback to guide the agent during training. The coarse mesh contains approximately 16 million degrees of freedom (DoF), while the fine mesh consists of around 418 million DoF. The computational cost per training step, considering the ten CFD simulations running in parallel, is $2{,}400$ CPU-hours. Each episode required three hours of wall-clock time on ten nodes of MareNostrum V, with each node equipped with four NVIDIA H100 GPUs and two Intel Xeon processors (80 CPU cores in total).

\begin{figure}[h]
	\begin{center}
	\includegraphics*[width=0.8\linewidth]{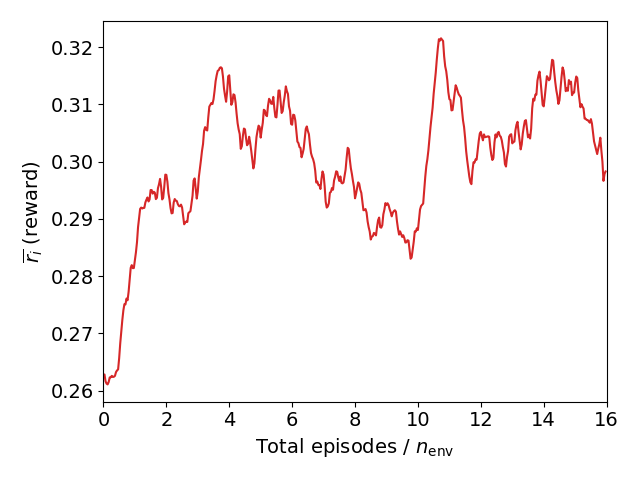}
	\caption{\label{fig:reward} Evolution of the local reward during the training with the coarse mesh.}
	\end{center}
\end{figure}

Fig. \ref{fig:reward} shows the evolution of the mean local reward ($\overline{r_i}$) over the final $5U_\infty / c$ of each episode. The rewards from the different environments that run in parallel  ($n_\text{env} = n_{\text{CFD}} \times n_{\text{MARL}}$) are sequentially appended in the plot. Then, by dividing the total number of episodes by the number of environments $n_\text{env}$, the number of training steps is obtained, as shown on the x-axis of the figure. In total, sixteen policy updates are performed and a certain level of learning can be detected, with the reward increasing during training.

\begin{figure}[h]
    \centering
    \begin{subfigure}[b]{0.95\linewidth}
        \includegraphics[width=\linewidth]{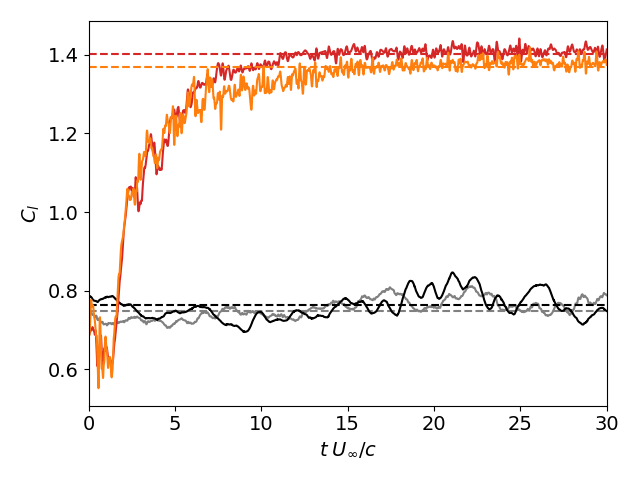}
        \caption{}
        \label{fig:Cl}
    \end{subfigure}
    \\
    \begin{subfigure}[b]{0.95\linewidth}
        \includegraphics[width=\linewidth]{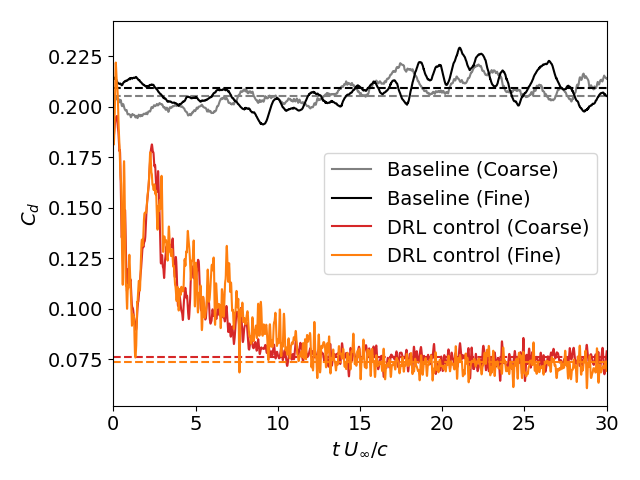}
        \caption{}
        \label{fig:Cd}
    \end{subfigure}
    \caption{\label{fig:ClCd}Temporal signals of (a) lift $C_l$ and (b) drag $C_d$ coefficients for both uncontrolled and DRL-controlled cases, evaluated using the coarse and fine meshes.}
\end{figure}

The lift and drag coefficient signals for both meshes in the uncontrolled scenario are shown in black (fine mesh) and grey (coarse mesh) in Fig. \ref{fig:ClCd}. It can be observed that both meshes produce very similar results in terms of flow statistics. To quantitatively assess the differences between meshes, Table \ref{tab:aerodynamic} presents a comparison of key aerodynamic metrics, including the mean lift coefficient $C_l$, mean drag coefficient $C_d$, the root-mean-square of lift fluctuations $C^{\prime}_{l, \mathrm{rms}}$ and the aerodynamic efficiency $E=C_l/C_d$. The differences between the two meshes are minimal: less than 2\% for the mean coefficients, and around 30\% for more sensitive parameters such as $C^{\prime}_{l,\mathrm{rms}}$, which was expected given that the coarse mesh does not fully capture the entire flow dynamics.

\begin{table*}[t]  
\centering
\caption{Summary of the main aerodynamic statistics for the baseline and DRL-controlled scenarios using both coarse and fine mesh resolutions.}
\label{tab:aerodynamic}
\begin{tabular}{cccccc}
\hline
Scenario & Mesh & $C_l$ & $C_d$ & $C^{\prime}_{l,\mathrm{rms}}$ & $E$ \\
\hline
\multirow{2}{*}{Baseline} & Coarse & 0.7480 & 0.2055 & 0.0226 & 3.640    \\
                          & Fine   & 0.7642 & 0.2095 & 0.0334 & 3.648    \\
\multirow{2}{*}{DRL-control} & Coarse & 1.4016 & 0.0763 & 0.0296 & 18.37    \\
                             & Fine   & 1.3685 & 0.0739 & 0.0205 & 18.52    \\
\hline
\end{tabular}
\end{table*}

Fig. \ref{fig:ClCd} also displays the $C_l$ and $C_d$ signals for the DRL-controlled scenarios, once the learned policy is evaluated without further exploration and for the two mesh refinement levels. An initial transient response is observed when the control is activated, lasting approximately $15U_\infty/c$. After this period, both signals settle into a statistically steady state that outperforms the baseline scenario. Once again, the results obtained using both meshes are highly consistent, supporting the use of the coarse mesh during training. As with the uncontrolled case, the main aerodynamic performance indicators for the controlled scenarios are summarised in Table \ref{tab:aerodynamic}. For the fine mesh, a lift enhancement of $\Delta C_l = 79\%$ is achieved, while the drag and the lift oscillations are reduced by $\Delta C_d = -65\%$ and $\Delta C^{\prime}_{l, \mathrm{rms}} = -39\%$, respectively, leading to a total efficiency improvement of $\Delta E = 408\%$. These results highlight the potential of DRL to discover highly effective active flow control strategies for fully separated flows.

\begin{figure}[h]
    \centering
    \begin{subfigure}[b]{0.95\linewidth}
        \includegraphics[width=\linewidth]{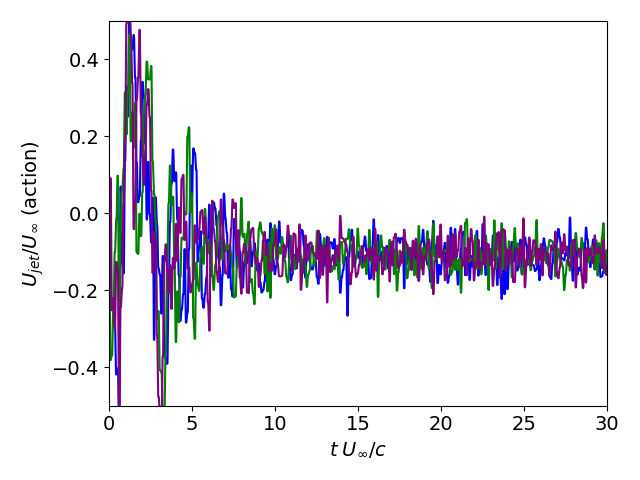}
        \caption{}
        \label{fig:actions_signals}
    \end{subfigure}
    \\
    \begin{subfigure}[b]{0.95\linewidth}
        \includegraphics[width=\linewidth]{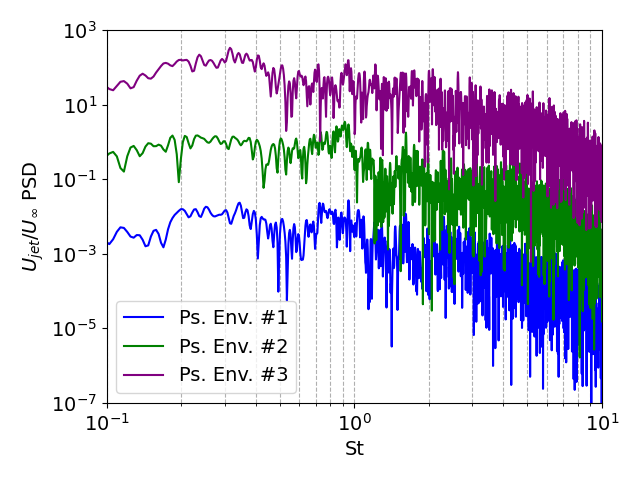}
        \caption{}
        \label{fig:actions_PSD}
    \end{subfigure}
    \caption{\label{fig:actions} (a) Jet velocity signals $U_{jet}/U_\infty$ for the front jets and (b) their corresponding energy spectra (vertically offset for clarity) for the fine mesh case.}
\end{figure}

The instantaneous actions applied by the agent (for the fine mesh) are shown in Fig. \ref{fig:actions_signals}, along with the corresponding frequency spectra of these signals in Fig. \ref{fig:actions_PSD}. For clarity, only the front jet signals are shown for the three MARL pseudo-environments, noting that the rear jet operates with an equal and opposite velocity. It is observed that, after the initial transient, the actions take negative values, indicating suction at the front jets and corresponding blowing at the rear jets. The mean velocity value of the front jets is around $\overline{U_{jet}}/U_\infty=-0.11$. This control strategy energises the flow along the leading edge of the wing, providing sufficient momentum to overcome the high curvature in this region and thereby prevent separation. In Fig. \ref{fig:actions_PSD}, although some discrete peaks appear near $\text{St} = 0.90$, no dominant frequencies are visible and the energy is broadly distributed across the turbulent spectrum. This indicates that the controller interacts with a wide range of coherent structures and continuously adapts to the instantaneous flow field. Thus, the effectiveness of the control strategy arises not only from the mean actuation but from its real-time responsiveness to unsteady flow dynamics, further illustrating the advantage of DRL and its closed-loop nature.

\begin{figure*}[t]
    \centering
    \begin{subfigure}[b]{0.4\linewidth}
        \includegraphics[width=\linewidth]{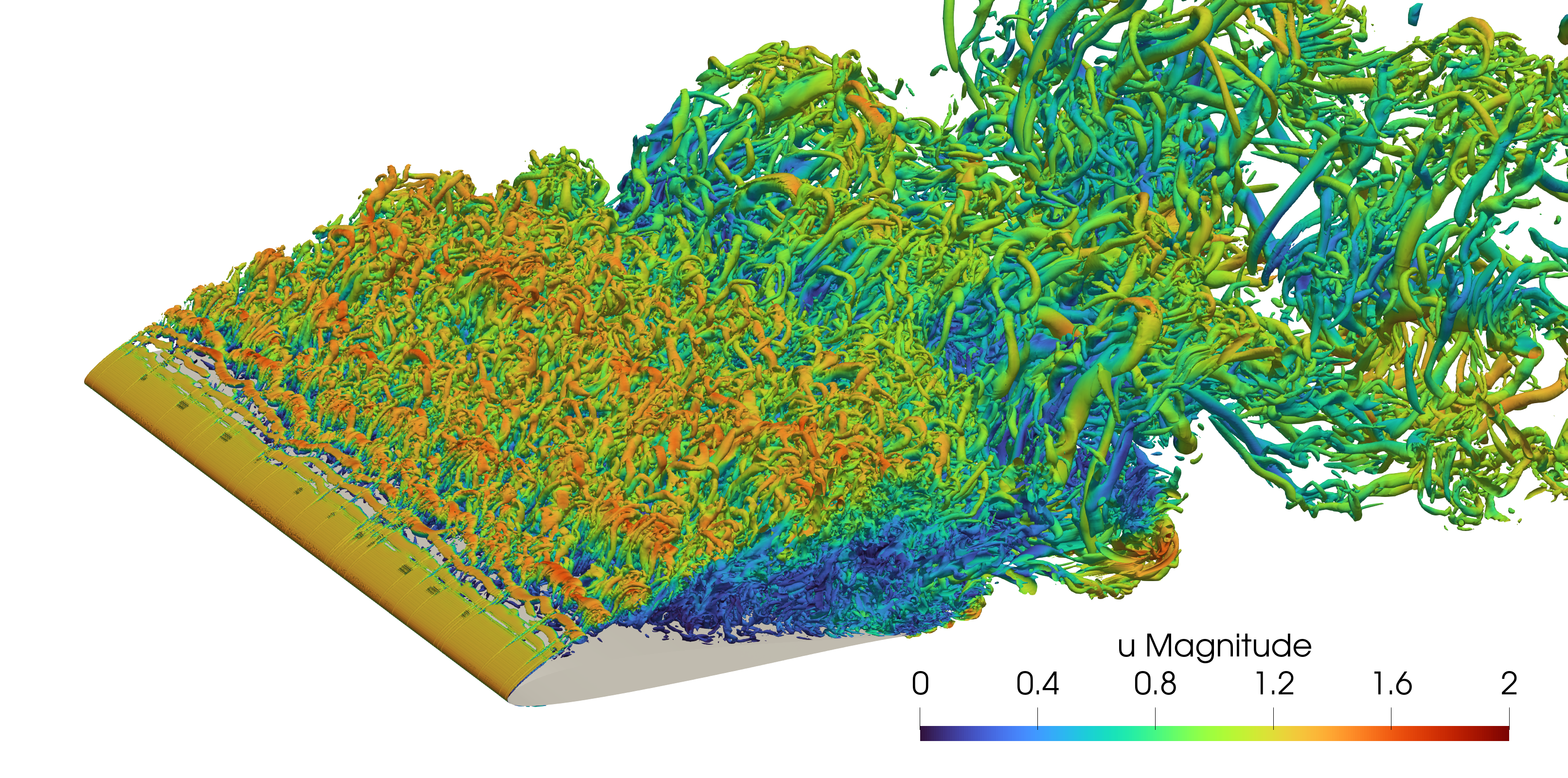}
        \caption{}
        \label{fig:Qs_base}
    \end{subfigure}
    \begin{subfigure}[b]{0.4\linewidth}
        \includegraphics[width=\linewidth]{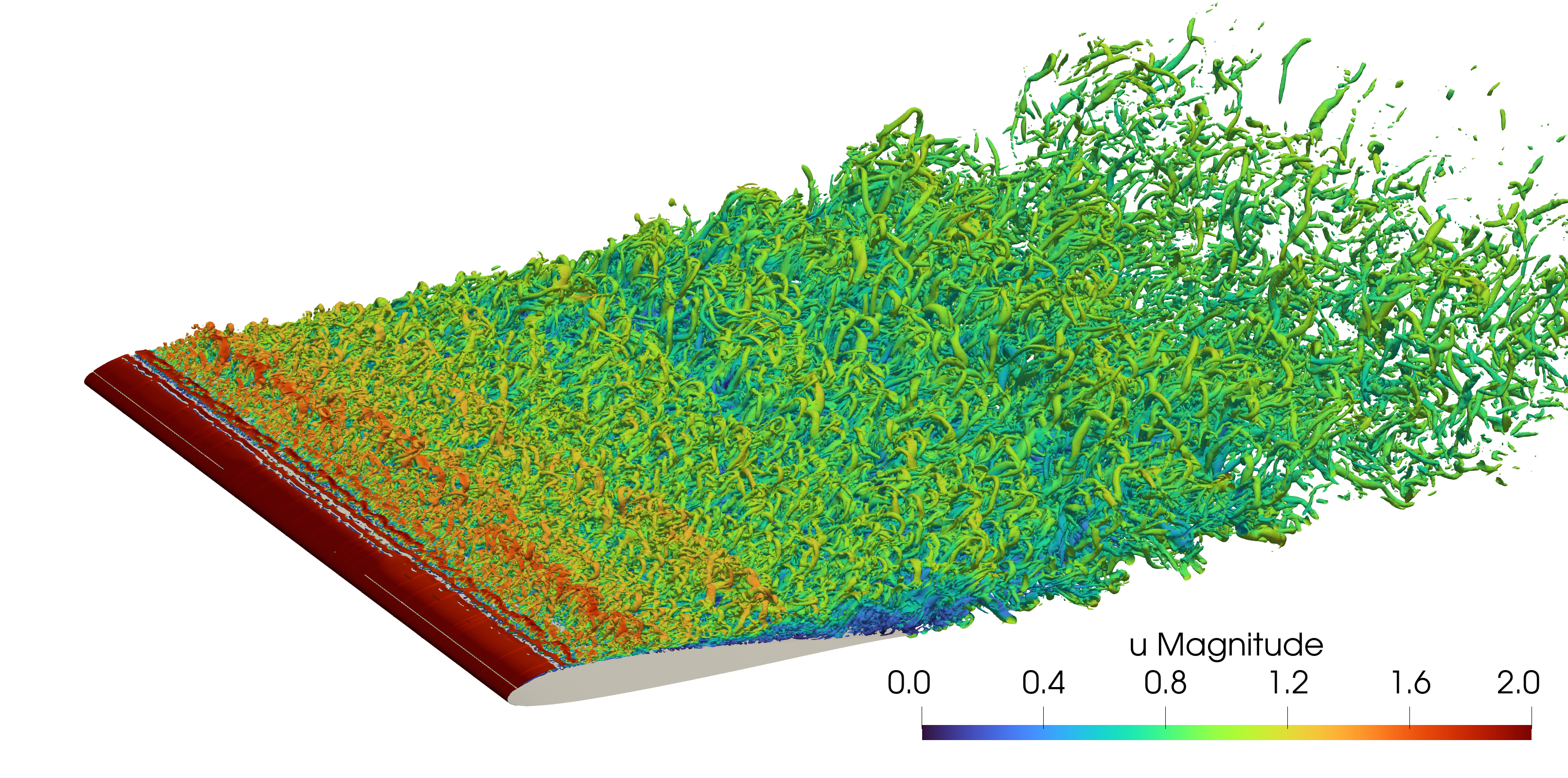}
        \caption{}
        \label{fig:Qs_control}
    \end{subfigure}
	\\
    \begin{subfigure}[b]{0.4\linewidth}
        \includegraphics[width=\linewidth]{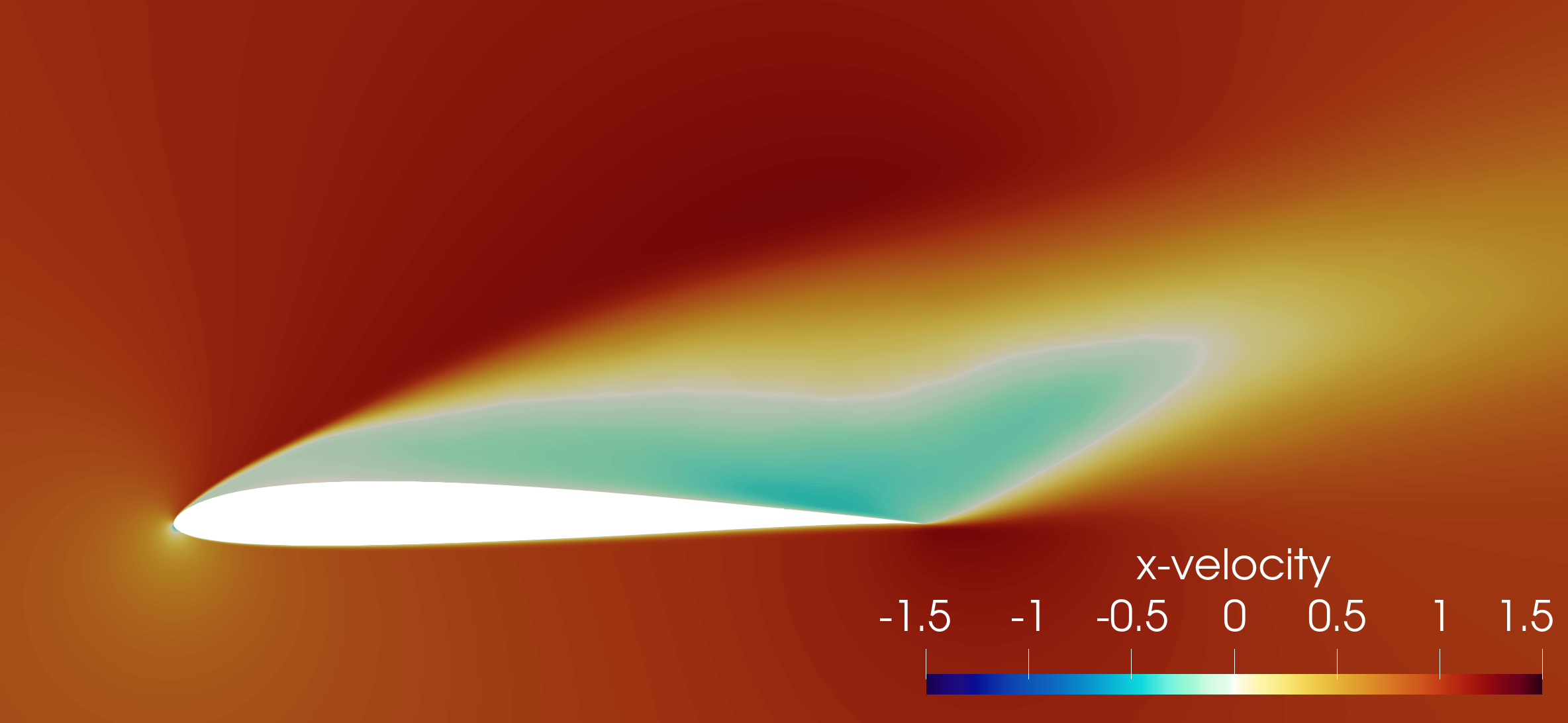}
        \caption{}
        \label{fig:Vel_base}
    \end{subfigure}
    \begin{subfigure}[b]{0.4\linewidth}
        \includegraphics[width=\linewidth]{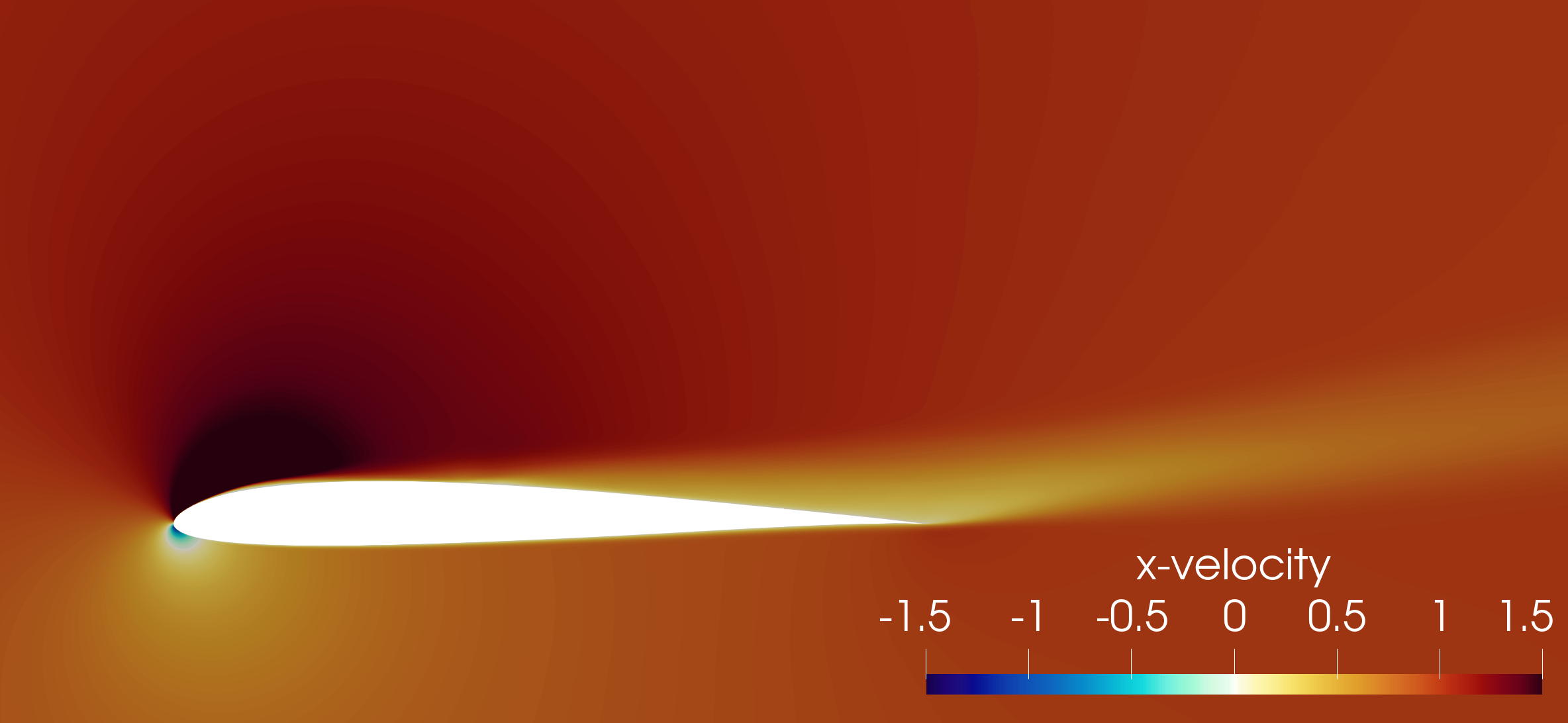}
        \caption{}
        \label{fig:Vel_control}
    \end{subfigure}
    \caption{\label{fig:3D_visualization}On the top row, iso-contours of the Q-criterion colored by the velocity magnitude $U/U_\infty$ for both the (a) uncontrolled and (b) DRL-controlled cases using the fine mesh. The bottom row presents the averaged x-velocity field $u/U_\infty$ on the $x-y$ plane for the (c) uncontrolled and (d) controlled scenarios, also computed with the fine mesh.}
\end{figure*}

Visualisations of the flow field are presented in Fig. \ref{fig:3D_visualization}. The top row shows vortical structures using iso-contours of the Q-criterion, while the bottom row displays the averaged streamwise velocity field on the $x-y$ plane. In the velocity contours, negative velocities are shown in blue, positive velocities in red, and the zero-velocity contour ($u=0$) is highlighted in white. This representation enables easy identification of recirculation regions within the flow. From this figure, the difference between the uncontrolled and DRL-controlled scenarios is clearly visible. In the baseline case, a large region of flow separation is observed, originating near the leading edge and extending along the entire chord of the wing (see Fig. \ref{fig:Qs_base} and Fig. \ref{fig:Vel_base}), due to the high angle of attack. When the control is activated, and after the transient period discussed before, the flow reattaches to the wing surface (see Fig. \ref{fig:Qs_control} and Fig. \ref{fig:Vel_control}), significantly improving aerodynamic performance, as reflected in the $C_l$ and $C_d$ signals previously shown. In the controlled scenario, some small and localised regions of negative x-velocity remain in the instantaneous field, although these disappear in the mean flow.

\section{Conclusions}
This study demonstrates the effectiveness of deep reinforcement learning (DRL) for active flow control (AFC) of fully separated flows at high angles of attack. Training on a computationally affordable coarse mesh allowed the agent to learn a control policy which was then successfully evaluated on a high-fidelity fine mesh. The DRL controller achieved a substantial lift enhancement (79\%), drag reduction (65\%), and improvement in aerodynamic efficiency (408\%) compared to the baseline. Flow visualizations confirmed significant reattachment of the separated shear layer, and the control actions revealed a physically interpretable strategy based on leading-edge suction and trailing-edge blowing. These results highlight the potential of DRL for discovering effective, data-driven control strategies in complex aerodynamic configurations.

\Acknowledgments
This research was partially funded by the Ministerio de Ciencia e Innovación of Spain (PID2023-150408OB-C21/C22). Simulations were supported by the Red Española de Supercomputación (IM-2024-2-0004 and IM-2024-3-0005) and EuroHPC JU (EHPC-REG-2024R01-038), granting access to MareNostrum V at the Barcelona Supercomputing Center. We also acknowledge AGAUR for supporting the LS/CFD (2021 SGR 00902) and TUAREG (2021 SGR 01051) research groups. R. M. thanks AGAUR for the FI-SDUR grant (2022 FISDU 00066) and R. V. acknowledges financial support from ERC grant no. 2021-CoG-101043998, DEEPCONTROL. Views and opinions expressed are however those of the author(s) only and do not necessarily reflect those of the European Union or the European Research Council. This work was initiated while R. V. was at KTH Royal Institute of Technology and completed during the transition to the University of Michigan.


\begin{References}
\item J.J. Cerutti, C. Sardu, G. Cafiero and G. Iuso (2020). Active Flow Control on a Square-Back Road Vehicle. {\it Fluids} 5, 55. https://doi.org/10.3390/fluids5020055
\item G. Minelli, S. Krajnović, B. Basara and B.R. Noack (2016). Numerical Investigation of Active Flow Control Around a Generic Truck A-Pillar. {\it Flow Turbul. Combust.} 97, 1235–1254. https://doi.org/10.1007/s10494-016-9760-3
\item I. Rodriguez, O. Lehmkuhl and R. Borrell (2020). Effects of the Actuation on the Boundary Layer of an Airfoil at Reynolds Number Re = 60000. {\it Flow Turbul. Combust.} 105, 607–626. https://doi.org/10.1007/s10494-020-00160-y
\item P. Garnier, J. Viquerat, J. Rabault, A. Larcher, A. Kuhnle and E. Hachem (2021). A review on deep reinforcement learning for fluid mechanics. {\it Comput. Fluids} 225, 104973. https://doi.org/10.1016/j.compfluid.2021.104973
\item J. Rabault, M. Kuchta, A. Jensen, U. Réglade and N. Cerardi (2019). Artificial neural networks trained through deep reinforcement learning discover control strategies for active flow control. {\it J. Fluid Mech.} 865:281–302. https://doi.org/10.1017/jfm.2019.62
\item J. Rabault and A. Kuhnle (2019). Accelerating deep reinforcement learning strategies of flow control through a multi-environment approach. {\it Phys. Fluids} 31 (9): 094105. https://doi.org/10.1063/1.5116415
\item P. Suárez, F. Álcantara-Ávila, J. Rabault, A. Miró, B. Font, O. Lehmkuhl and R. Vinuesa (2025a). Flow control of three-dimensional cylinders transitioning to turbulence via multi-agent reinforcement learning. {\it Commun. Eng.} 4, 113. https://doi.org/10.1038/s44172-025-00446-x
\item P. Suárez, F. Álcantara-Ávila, A. Miró, J. Rabault, B. Font, O. Lehmkuhl and R. Vinuesa (2025b). Active flow control for drag reduction through multi-agent reinforcement learning on a turbulent cylinder at $Re_D=3900$. {\it Flow Turbul. Combust}. https://doi.org/10.1007/s10494-025-00642-x
\item L. Guastoni, J. Rabault, P. Schlatter, H. Azizpour and R. Vinuesa (2023). Deep reinforcement learning for turbulent drag reduction in channel flows. {\it Eur. Phys. J. E} 46, 27. https://doi.org/10.1140/epje/s10189-023-00285-8
\item J. Vasanth, J. Rabault, F. Alcántara-Ávila, M. Mortensen and R. Vinuesa (2024). Multi-agent Reinforcement Learning for the Control of Three-Dimensional Rayleigh–Bénard Convection. {\it Flow Turbul. Combust.} https://doi.org/10.1007/s10494-024-00619-2
\item Y. Z. Wang, Y. F. Mei, N. Aubry, Z. Chen, P. Wu and W. T. Wu (2022). Deep reinforcement learning based synthetic jet control on disturbed flow over airfoil. {\it Phys. Fluids} 34(3): 033606. https://doi.org/10.1063/5.0080922
\item X. Garcia, A. Miró, P. Suárez, F. Alcántara-Ávila, J. Rabault, B. Font, O. Lehmkuhl and R. Vinuesa (2025). Deep-reinforcement-learning-based separation control in a two-dimensional airfoil. {\it Int. J. Heat Fluid Flow} 116: 109913. https://doi.org/10.1016/j.ijheatfluidflow.2025.109913
\item B. Font, F. Alcántara-Ávila, J. Rabault, R. Vinuesa and O. Lehmkuhl (2025). Deep reinforcement learning for active flow control in a turbulent separation bubble. {\it Nat. Commun.} 16, 1422. https://doi.org/10.1038/s41467-025-56408-6
\item R. Montalà, B. Font, P. Suárez, J. Rabault, O. Lehmkuhl, R. Vinuesa and I. Rodríguez (2024). Towards Active Flow Control Strategies Through Deep Reinforcement Learning, in: 9th ECCOMAS, 3rd-7th June, Lisbon (Portugal). https://doi.org/10.23967/eccomas.2024.115
\item R. Montalà, B. Font, P. Suárez, J. Rabault, O. Lehmkuhl, R. Vinuesa and I. Rodriguez (2025). Deep Reinforcement Learning for Active Flow Control around a Three-Dimensional Flow-Separated Wing at Re = 1000, in: 1st International Symposium on AI and Fluid Mechanics, 27th-30th May, Chania (Greece).
\item L. Gasparino, F. Spiga and O. Lehmkuhl (2024). SOD2D: A GPU-enabled
spectral finite elements method for compressible scale-resolving
simulations. {\it Comput. Phys. Commun.} 297, 109067. https://doi.org/10.1016/j.cpc.2023.109067
\item U. Piomelli, A. Rouhi and B. J. Geurts (2015). A grid-independent length scale for large-eddy simulations. {\it J. Fluid Mech.} 766:499-527. https://doi.org/10.1017/jfm.2015.29
\item V. Belus, J. Rabault, J. Viquerat, Z. Che, E. Hachem and U. Reglade (2019). Exploiting locality and translational invariance to design effective deep reinforcement learning control of the 1-dimensional unstable falling liquid film. {\it AIP Adv.} 9 (12): 125014. https://doi.org/10.1063/1.5132378
\item S. Guadarrama, A. Korattikara, O. Ramirez, P. Castro, E. Holly, S. Fishman, K. Wang, E. Gonina, N. Wu, E. Kokiopoulou, L. Sbaiz, J. Smith, G. Bartók, J. Berent, C. Harris, V. Vanhoucke and E. Brevdo (2018). TF-Agents: A library for Reinforcement Learning in TensorFlow. https://github.com/tensorflow/agents
\item S. Partee, M. Ellis, A. Rigazzi, A. E. Shao, S. Bachman, G. Marques and B. Robbins (2022). Using machine learning at scale in numerical simulations with SmartSim: An Applications to ocean cimate modeling. {\it J. Comput. Sci.} 62, 101707. https://doi.org/10.1016/j.jocs.2022.101707
\item J. Schulman, F. Wolski, P. Dhariwal, A. Radford and O. Klimov (2017). Proximal policy optimization algorithms. {\it Preprint at} https://doi.org/
10.48550/arXiv.1707.06347
\end{References}

\end{document}